# Application of single metal Au and Ag/Si$_3$N$_4$/Au plasmonic sensors for glucose refractive index measurements

H. Akafzade[1], N. Hozhabri[2], and S.C. Sharma[1,*]

[1]Department of Physics, University of Texas at Arlington, Arlington, TX 76019

[2]Nanotechnology Research Center, Shimadzu Institute, University of Texas at Arlington, Arlington, TX 76019

[1]To whom correspondence should be addressed (sharma@uta.edu)

## ABSTRACT

We have employed single gold layer metal and multilayer Ag/Si$_3$N$_4$/Au plasmonic sensors to monitor changes in the refraction index of glucose/water solutions. Our results show that the sensor can resolve glucose refractive index down to 1% concentration. In addition, we have simulated the strength of the surface plasmon resonance evanescent electric fields by using the finite difference method of COMSOL. The electric field strength for the Ag/Si$_3$N$_4$/Au sensor is much stronger than that for single gold layer SPR sensor.

## 1. Introduction:

Surface Plasmon Polaritons (SPPs) are electromagnetic excitations related to surface charge density oscillation propagating along a metal/dielectric interface. Since the discovery of SPPs over seven decades ago, their applications have led to the development of various Surface Plasmonic Resonance (SPR) sensors with applications in bioscience, engineering, chemical, and physical applications.[1-6] The simplest surface plasmon resonance (SPR) sensor is made of a thin layer of metal film such as silver or gold on glass. Other forms of SPR sensors are stack of two metals such as silver and gold and more complicated nanostructures consisting of metal/dielectric/metal sensors.[7-10] Each type of these sensors has advantages and disadvantages depending on the application and ease of manufacturability. For example, the single metal Ag SPR sensor is the simplest and easiest to manufacture, however, it has a short life expectancy. In case of silver, the oxidation of silver results in significant degradation of the sensor in short period of time.[11] On the other hand, stack of three layers SPR sensors such as Ag/Si$_3$N$_4$/Au last longer, however, its sensitivity is lower than that of a silver sensor. In previous works, we have studied single metal, two metal and metal/dielectric/metal sensors.[8-10, 12] The condition for the excitation of Surface Plasmon Resonance for a single metal film based on $n_p sin\theta_{spr} = \sqrt{\frac{\varepsilon_m n_d^2}{\varepsilon_m + n_d^2}}$, where $n_p$ is the index of prism, $\varepsilon_m$ is real part of dielectric constant of metal, and $n_d$ is the ambient index of refraction. The sensitivity of a SPR sensor is defined as the derivative of the SPR angle with respect to index of refraction of ambient or $S_n = d\theta_{spr}/dn_d$

## 2. Experimental Details:

In this work, we have fabricated two sets of sensors of Au and Ag/Si$_3$N$_4$/Au structures in a class-100 clean room followed by SPR measurements for the performance and sensitivity of these sensors. Single metal Ag SPR sensor acts as the base for comparison due to the fact that it has the highest sensitivity. Prior to device fabrication, each device was simulated with COMSOL software to determine the thickness of each layer for optimum SPR. Optimized structures are 50nm for single metal gold and 40nm/186nm/22nm for the Ag/Nitride/Au sensor. Metal depositions were made in an AJA thermal evaporator at base pressure of $1.2 \times 10^{-7}$ torr. Prior to fabrication, the glass substrates were cleaned in piranha solution followed by DI water rinse and nitrogen gas dry. Substrates were annealed at 70º C for over ten (10) minutes for degassing. Gold and silver metal films were fabricated at deposition rate of ~ 0.5 Å per second at room temperature. In case of multilayer stack sensor, samples were immediately transferred to an AJA sputtering deposition equipment for silicon nitride deposition at a base pressure of mid $10^{-8}$ torr. Post nitride deposition, stack sensor was transferred back to AJA thermal evaporator for gold deposition. Condition for gold deposition was as before. To determine the actual thickness films, a photoresist coated and patterned silicon sample was mounted next to SPR sample at each stage of fabrication in deposition equipment. The patterned silicon sample was then processed in acetone for photoresist removal followed by thickness measurement on a KLA-TENCOR P6 profilometer. Our experimental set up for SPR measurements has been described elsewhere.[7, 8] It uses standard silicon detectors for measurements of the relative intensities of the incident and reflected light from 12-mW He-Ne laser. A set of polarizers is used to obtain *p*-and *s*-polarized beams. For the measurements reported here, the standard thigh-index prism is replaced by a half-cylindrical lens of 1.5 refractive index. The use of the half-cylindrical lens of 1.5 refractive index allows angular scans over a wider range of angles. The sensor is coupled by refractive-index matching fluid to the surface of the lens, which is mounted on a $(\theta, 2\theta)$ rotating table for angular scans. A LabView software is used for data collection. The half-cylindrical lens arrangement is shown in figure 1.

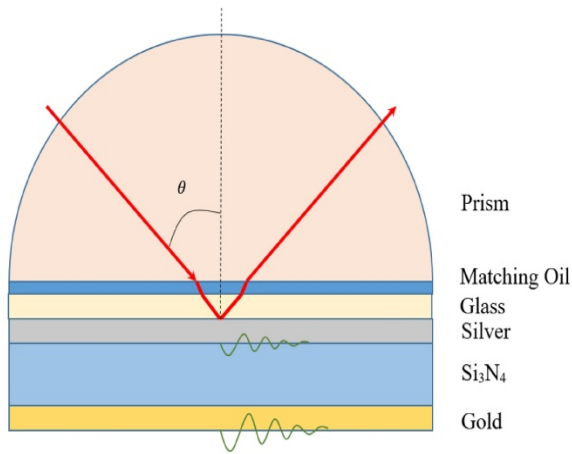

Figure 1. Half-cylindrical lens arrangement showing coupling to glass with refractive index matching oil and Ag/Si$_3$N$_4$/Au layers

## 3. Results and discussions:

First of all, we evaluate the influence of the refractive index of the dielectric (the analyte, e. g., glucose water solution) on the SPR angle and sensitivity of the sensor. Figure 2 presents these results for glass prism ($n_p$ = 1.496) and single metal (gold) SPR sensor. The SPR angle changes from about $40^0$ to $90^0$ for refractive index values from 1.0 to 1.37. The sensor sensitivity is about 60 degrees for air and rapidly increases after index of 1.2 and for example reaches 220 degrees per refractive index unit (RIU), for index of approximately 1.345. The upper level index of ambient material that can be determined, as well as sensitivity of detection, is a function of index of prim, as evident in Figure-2. In case of gold sensor, the theoretical sensitivity is ~$180^0$. The green line represents distilled water.

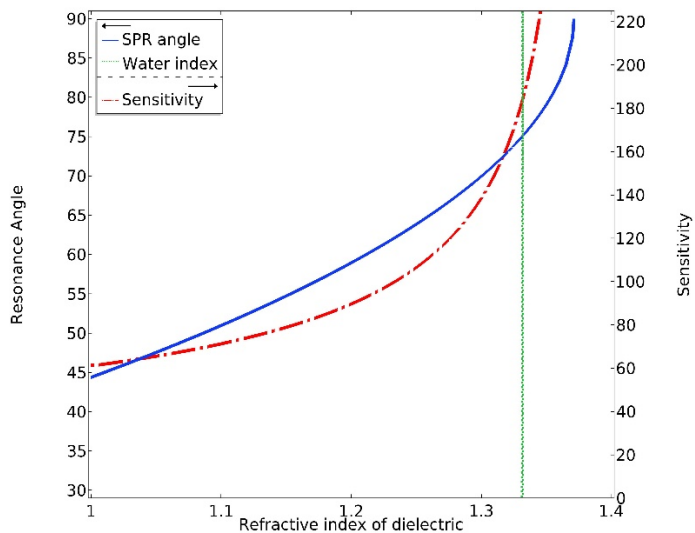

Figure 2. The SPR angle sensor sensitivity as a function of index of the dielectric.

### 3.1. Single layer gold SPR sensor:

We measured SPR response for distilled water and glucose for glucose concentrations of 1%, 2%, 5%, 10% and 20%. Figure 3 shows partial measurement results for SPR data for glucose concentration of 10% and 20% and simulation results. The plot of the SPR angle of incident as a function of water and glucose index of refraction is given in Figure 4. Data show almost a linear relation between SPR angle and the index of refraction that is concentration dependent. The slope of the linear fit to the data provides the sensitivity for the single metal gold SPR sensor, $167^0$. This is less than the simulated value of $180^0$. The difference in the expected and measured value is attributed to several factors including uncertainty of measured SPR incident angle, the deviation of angle at interface of glass and prism and matching oil that can result in small but effective parameter to alter the incident angle. Based on these measurements, the lowest concentration that can be resolved by a single layer gold SPR sensor is 1% per volume.

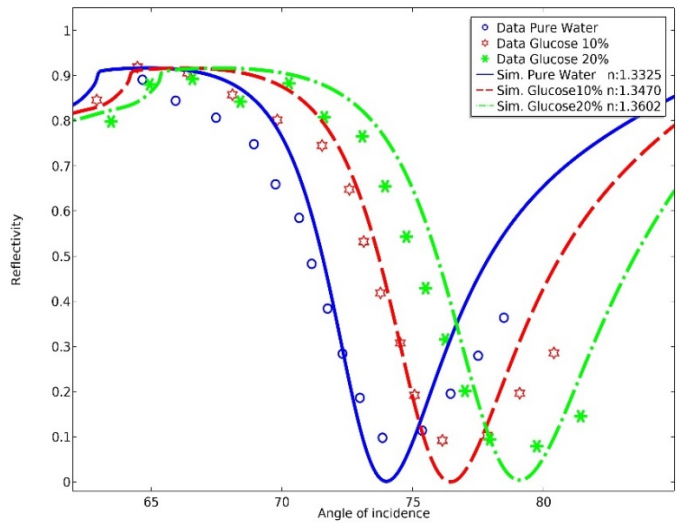

Figure 3. Reflectivity as a function of the angle of incidence for incident laser for single gold layer

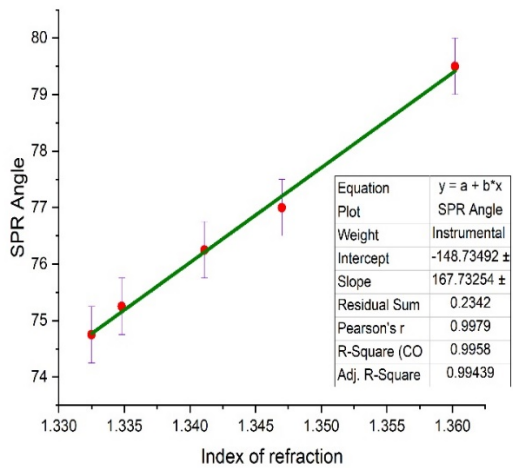

Figure 4. SPR angle vs. index of refraction of various glucose concentrations for single gold layer sensor

### 3.2. Multilayer Ag/Si$_3$N$_4$/Au SPR Sensor:

The second sensor we tested in this work is Ag/Nitride/Au SPR Ag/Si$_3$N$_4$/Au Sensor. Again, thickness of each layer was simulated prior to fabrication. We measured the thickness of each step after deposition and total reflectivity was simulated for the remaining layers. At each layer, after measurement, adjustment was made to parameters for simulation to optimize the next layers accordingly. The final thicknesses are 49nm/186nm/21nm for SPR signal of over 90% reflectivity. We measured the SPR signal for solutions of glucose concentrations of 2%, 5%, 10%, 20%, and 40% in addition to distilled water (zero glucose concentration). Again, the index of refraction for each concentration was determined by matching the SPR simulation to SPR data for each solution. The sensitivity of the sensor as before was determined from the plot of SPR angle vs. Index of refraction of the solution and it's shown in Figure 6. The sensitivity of this sensor is measured to be ~ 91º. Lower sensitivity is consistent with inability of measurement for 1% glucose concentration. The lowest concentration that we could resolve with this sensor is 2%.

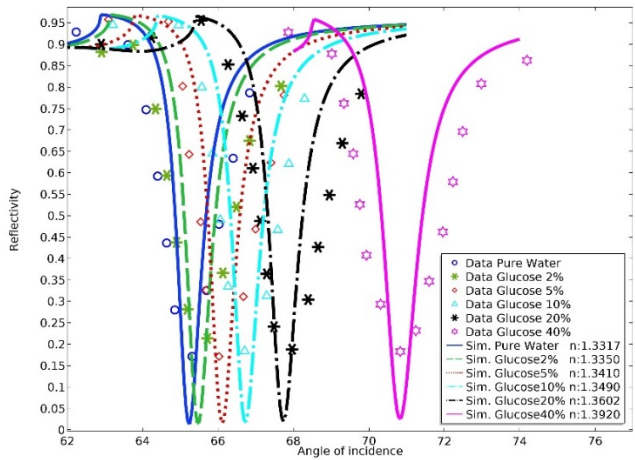

Figure 5. Reflectivity as a function of angle of incidence for varying relative concentrations of glucose for multilayer sensor

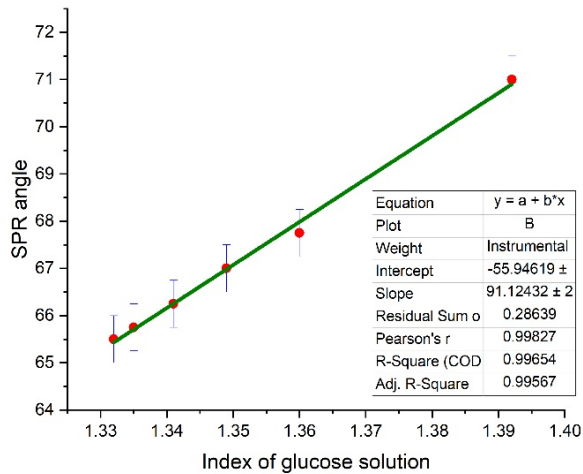

Figure 6. SPR angle of Ag/Si3N4/Au sensor vs. index of refraction of various glucose concentrations

In addition, we extracted the glucose index of refraction at tested glucose concentration by matching the simulation parameters to experimental SPR data. The result is provided in Figure 7.

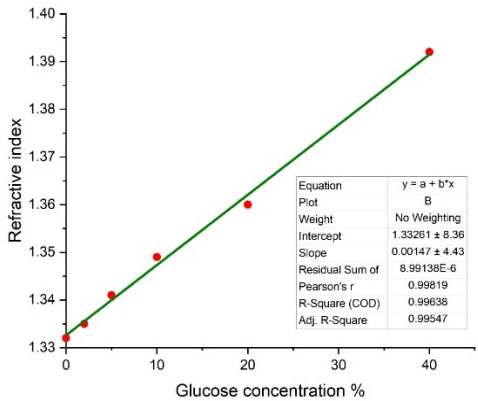

Figure 7. Glucose index of refraction as function of glucose concentration of zero (distilled water) to 40%. Solid line is linear fit to the data.

## 3.3. Electric Fields

To further investigate these sensors, we simulated the electric fields for both sensors based on results of the layer thicknesses. In case of the first sensor, the figures 8-9 show the strength of the z-component (normal to the plane of the sensor) electric fields for single metal gold and also for Ag/Si$_3$N$_4$/Au sensors. A comparison of the electric fields of the gold and Ag/Nitride/Au sensors shows that electric field for the Ag/Nitride/Au sensor is at least 1.5 times stronger with about 300nm higher decay length. It amounts to detection capability up to much higher range from the surface.

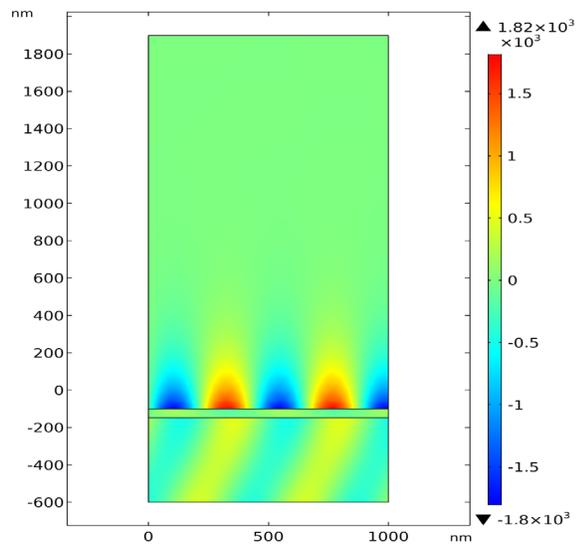

Figure 8: Electric filed in z-direction for single gold metal sensor in contact with water at SPR angle

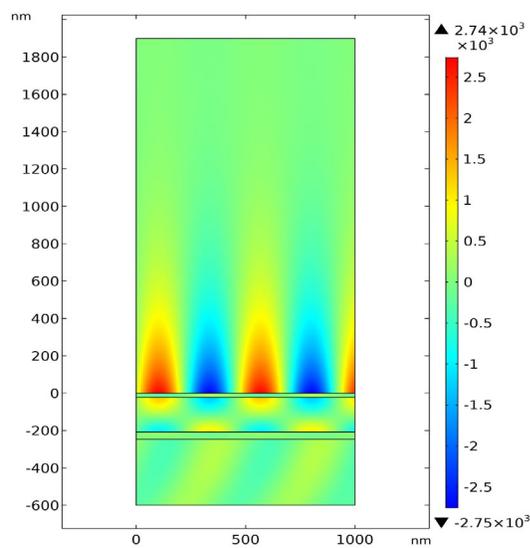

Figure 9: Electric filed in z-direction for Ag/Si3N4/Au in contact with water at SPR angle.
77

## 4. Conclusions

In conclusion, we have applied single metal gold and also Ag/Si$_3$N$_4$/Au SPR sensors to determine the glucose refractive index for glucose concentrations, varying from 1% to 40 %. To our knowledge, the sensitivity of the gold sensor to glucose concentration is higher than other known sensors. [13-15] However, the electric field of the Ag/Si$_3$N$_4$/Au sensor and its range are 1.5 times higher than the Au sensor with longer range of detection. By improving the setup and having a rotation stage that can read angle with higher precisions, we can significantly increase the performance of this sensor and detect the change in concentration with much higher accuracy than it is now. For example if a rotation stage can distinguish the change in angle as much as 0.016 degrees, which is commercially available, then by this sensor we can potentially detect the change in refractive index as much as 0.0001, which is suitable for biology and medical purposes.